# A Corpus-based Evaluation of Lexical Components of a Domain-specific Text to Knowledge Mapping Prototype

Rushdi Shams[1] and Adel Elsayed[2]

Department of Computer Science and Engineering, Khulna University of Engineering & Technology (KUET), Bangladesh[1]
M3C Research Lab, the University of Bolton, United Kingdom[2]

*Abstract* — The aim of this paper is to evaluate the lexical components of a Text to Knowledge Mapping (TKM) prototype. The prototype is domain-specific, the purpose of which is to map instructional text onto a knowledge domain. The context of the knowledge domain of the prototype is physics, specifically DC electrical circuits. During development, the prototype has been tested with a limited data set from the domain. The prototype now reached a stage where it needs to be evaluated with a representative linguistic data set called corpus. A corpus is a collection of text drawn from typical sources which can be used as a test data set to evaluate NLP systems. As there is no available corpus for the domain, we developed a representative corpus and annotated it with linguistic information. The evaluation of the prototype considers one of its two main components- lexical knowledge base. With the corpus, the evaluation enriches the lexical knowledge resources like vocabulary and grammar structure. This leads the prototype to parse a reasonable amount of sentences in the corpus.

## I. INTRODUCTION

Text to Knowledge Mapping (TKM) Prototype [1] is a domain-specific NLP system, the purpose of which is to parse instructional text and to model it with its pre-defined ontology. During development, the prototype has been tested with a limited data set from the domain Physics instructional text on DC electrical circuit. The prototype now reached a stage where its lexical components need to be evaluated with a representative linguistic data set, a corpus- a collection of text drawn from typical sources. Information retrieved during parsing contributes to map and model domain-specific text on its knowledge domain. Therefore, the usability of the TKM prototype as a specialized knowledge representation tool for the domain depends on the evaluation of its lexical components.

An important precondition to evaluate NLP systems is the availability of a suitable set of language data called corpus as test and reference material [2]. With an extensive web-based search, we did not find any corpus for the domain DC electrical circuit. Therefore, we need to develop a representative corpus to evaluate the lexical components of the prototype because a representative corpus reflects the way language is used in the domain [3]. A usable corpus requires various annotations according to the scope and type of evaluation. As we intend to evaluate the lexical components of the TKM prototype, the corpus should be annotated with linguistic information like Parts of Speech (POS) tagging, phrasal structure annotations, and stem word tagging. These annotations can lead us to adjust the lexical components of the prototype according to the qualitative and quantitative layers [1] [4] of its knowledge model.

In this paper, we proposed a stochastic development procedure of a domain-specific representative corpus that is used to evaluate the lexical components of the TKM prototype. We also represented detail procedure of corpus-based evaluation of an NLP system- that includes enriching the lexicon and morphological database, testing the parsing ability of the prototype, and the adjustment of the lexical components according to the linguistic information in the corpus.

The remainder of this paper is organized as follows. In Section II, corpus-based evaluations of various NLP systems have been discussed. Section III describes the proposed procedure of representative corpus development and annotations. Section IV describes the evaluation of lexical components of the TKM prototype such as the vocabulary and grammar structure. Section V concludes the paper.

## II. RELATED WORK

A text based domain-specific NLP system can be evaluated according to the type, context or discourse of text from the domain although no established agreement has been developed on test sets and training sets [5]. Corpus is not restricted today only for researches on linguistics [6]; it is now becoming the principal resource to evaluate such domain-specific NLP systems. Many NLP systems like Saarbrucker Message Extraction System (SMES) [8] have been tested with a corpus as proper evaluation depends on a representative test set of data like corpus [7]. Corpus contains structured and variable but representative text. A corpus is said representative if the findings from it can be generalized to language or a particular aspect of language as a whole [3]. Corpus-based evaluations like MORPHIX [13] and MORPHIX++ [7] showed that the evaluation with a representative corpus results in proper adjustments. MORPHIX++ was tested with a corpus and systematic inspection revealed some necessary adjustments like missing lexical entries, discrepant morphology incomplete or erroneous single words.

Now-a-days, many web corpora are present and researchers successfully utilize web data including machine translation [9], prepositional phrase attachment [10] and other anaphora resolution [11] to evaluate NLP systems. One problem in evaluating domain-specific NLP systems with a web corpus is the discrepancy between the domain of the corpus and the domain of the system [14]. Having the ease of access [15], the web corpora development tools like BootCaT [16] and WebBootCaT [17] are used world wide to develop corpus using web data.

NLP systems use either pre-defined or customized grammar rules. For instance, the lexical components of the TKM prototype use Combinatory Categorical Grammar (CCG) [12]. The prototype follows some specific clausal and phrasal structures according to CCG. As it follows a particular grammar, we need to adjust the grammar and phrasal structures according to the structures of text from the domain. For example, TKM prototype, on its early test, was able to parse simple sentences only. This becomes a drawback if majority of text in the domain is written in compound and complex sentences. Therefore, necessary adjustment on CCG can let the prototype parse compound and complex sentences as well. In addition, NLP systems may recognize specific clausal and phrasal structure which maybe absent in domain-specific text. For example, if an NLP system uses grammars that handle one subject and one object, both parsing and knowledge extraction from domain-specific text becomes difficult if majority of the text contains more than one subject and one object. These linguistic properties of domain-specific text bring in the issue of adjustment. The lexicographical resources of such systems can be increased by analyzing linguistic patterns in domain-specific corpus. Statistical data like frequency of words, number of simple, complex or compound sentences, number of subject and object present in the sentences assist to adjust the lexical components of the systems. The grammar structure MORPHIX++ supported was not efficient in its early days. It was adjusted and extended according to the corpus used as its test suite.

III. CORPUS DEVELOPMENT

In this section, we will discuss regarding the development approach of a domain-specific corpus, proof or its representativeness, and its annotation procedure.

*A. Development Approach*

As we did not find any corpus for the domain DC electrical circuit with extensive web searches, we initiated WebBootCaT to develop a representative corpus. We developed five corpora using the WebBootCaT and analyzed them by comparing the number of distinct domain-specific terms and number of distinct words present. The significant difference between these two numbers and inconsistency on the size of the corpus in Fig. 1 state that web-based tools are not usable to develop domain-specific corpora, but maybe useful for developing domain independent corpora.

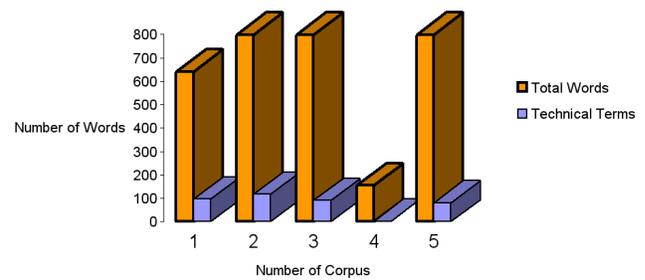

Fig. 1. Inconsistency of WebBootCaT to develop a domain-specific corpus

WebBootCaT works on the basis of searching the web according t-o user defined keywords called seeds. Researches like [15] showed that search engines like Google provide 67 search results for Speculater and about 82,000 search results for Speculator?- which vividly shows the reason behind inconsistent nature of web-based corpus developing tools. Therefore we decided to develop the corpus manually and collected text from 141 web resources containing 1,029 sentences and 18,834 words. During the development, we left the non textual information (e.g., equations and diagrams) as the TKM prototype operates only on text.

*B. Representativeness of the Corpus*

The representativeness of the corpus can be justified with a notion of saturation or closure described by [18]. At the lexical level, saturation can be tested by dividing the corpus into equal sections in terms of number of words or any other parameters. If another section of the identical size is added, the number of new items in the new section should be approximately the same as in other sections [19].

To find out the representativeness for the corpus, it has been segmented into 15 samples. Each sample is comprised of 1,267 words on average. We plotted the cumulative frequency of the most frequent technical terms in the samples.

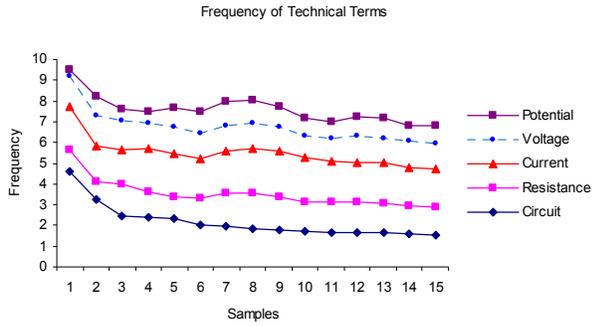

Fig. 2. Representativeness of the corpus with domain-specific technical terms

Fig. 2 depicts that the presence of the domain-specific technical terms becomes stationary after a few samples. This is one of the criteria showing the representativeness of the corpus. After a certain point, no matter how much text we add to the corpus, the frequencies of the terms are becoming stationary.

Similarly, we counted the frequency of non-technical words in the corpus and grouped them according to their parts of speech. Statistics on verbs, prepositions and co-ordinators in Fig. 3 (a), 3 (b) and 3 (c) respectively show that the corpus has been saturated after sample 11.

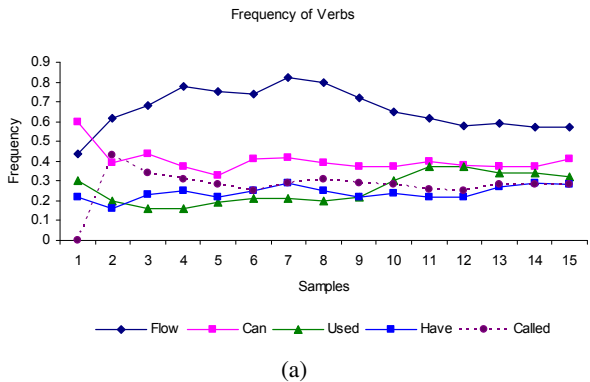

(a)

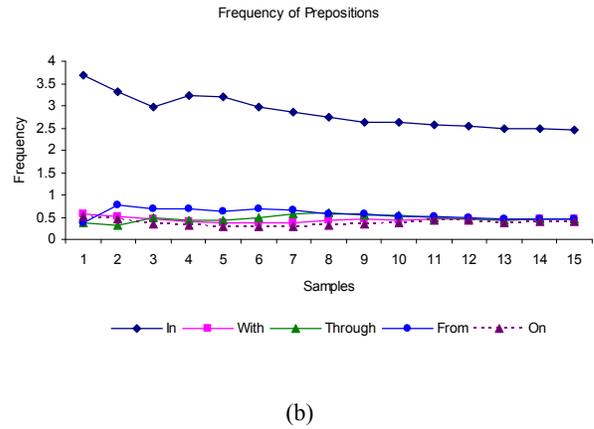

(b)

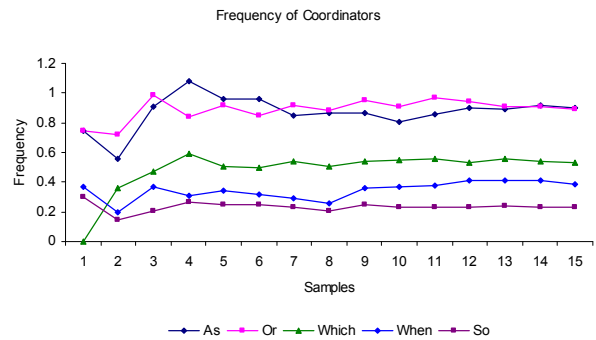

(c)

Fig. 3. Representativeness of the corpus with (a) verbs, (b) prepositions and (c) coordinators

We also counted the frequency of types of sentences in the corpus. As the domain contains instructional text and most of which are simple sentences, it needs to be reflected on the corpus as well. Fig. 4 shows that majority of the text is simple sentence (in percentage).

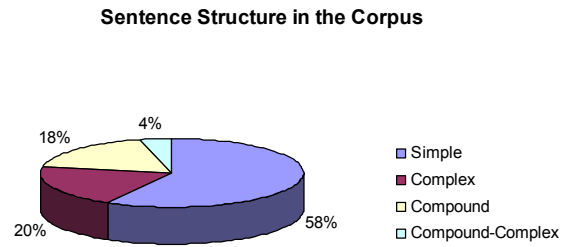

Fig. 4. Sentence structure in the corpus

## C. Corpus Annotation

To annotate the corpus with POS tags, Cognitive Computation Group POS tagger [20] has been used as it works on the basis of learning techniques like Sparse Techniques on Winnows (SNOW). The corpus is annotated with nine parts of speech include noun, pronoun, verb, adverb, adjective, preposition, coordinator, determiner, and modal. The phrasal structure of the corpus has been annotated by the slash-notation grammar rules defined by CCG. We developed seven XML tags including URL, date, sentence, word, stem, pos, and category and created an XML version of the corpus.

## IV. EVALUATION OF LEXICAL COMPONENTS

The evaluation of vocabulary and the grammar structure of the prototype are illustrated in this section. This section also refers to the efficiency in parsing and richness of lexical entries of the prototype.

### A. Evaluation of Vocabulary

The lexicon of the prototype is mapped on the unique words of the corpus. The words present both in the morphology and in corpus are called the vocabulary of the prototype. Initially, only five percent of the vocabulary was covered by the prototype (Table I).

TABLE I
PRELIMINARY VOCABULARY COVERAGE OF THE TKM PROTOTYPE

| Words in Morphology and in Corpus | Unique Words in the Corpus | Vocabulary Covered |
|---|---|---|
| 101 | 1,902 | 5% |

MORPHIX++, a second generation NLP system, covered 91 percent of word in the corpus developed to evaluate it. The reason behind this difference is the augmentation of the vocabulary of MORPHIX++ ran parallel with the development of the system. In contrast, the main focus during the development of the TKM prototype was to develop an operational system first rather than increasing its vocabulary.

We used the POS tags of the corpus to populate the lexicon. We retrieved every distinct word for each distinct POS from the corpus and we simply added it if that word was absent in the lexicon. The number of added entries into the lexicon is shown in Table II. On completion of the process, the vocabulary of the prototype covers 90 percent of the corpus (Table III).

TABLE II
AUGMENTATION OF LEXICAL ENTRIES IN THE TKM PROTOTYPE

| POS | Augmented Entries |
|---|---|
| Determiner | 19 |
| Coordinator | 5 |
| Noun and Pronoun | 2,094 |
| Adjective | 364 |
| Preposition | 71 |
| Adverb | 177 |
| Verb | 264 |

TABLE III
VOCABULARY OF THE TKM PROTOTYPE

| Words in Morphology and in Corpus | Unique Words in the Corpus | Vocabulary Covered |
|---|---|---|
| 1,783 | 1,902 | 90% |

### B. Evaluation of Grammar Structure

The TKM prototype struggles to parse *modals* or *auxiliary verb* because CCG does not provide any specification to categorize modals into finite and non-finite [21]. We defined grammar formalisms for modals and adjusted the lexicon that increased the ability of the prototype to parse modals.

CCG does not have any mechanism for phrasal structures like *adjective–adjective–noun* although researches showed that numerous adjectives can be placed before a noun [22]. Moreover, CCG treats adjectives as qualifiers of noun (*np/n* or *n/n*) but in the sentence *when the lamps are connected in series, the more the lamps, the more dimmer they get*, *dimmer* is not particularly preceding any noun or pronoun. Except the regular adjectives, we defined grammar formalisms for noun equivalents (e.g., two *common* types of circuits), participle equivalent (e.g., the *connected* wire), gerund equivalents (e.g., the *conducting* material), and adverb equivalents (e.g., the *above* circuit is series circuit) of adjectives that increased the rate of parsing adjectives.

CCG has specification for pronoun by treating it as a noun without any anaphoric resolves [23]. For example, *it is placed in parallel*- CCG can parse this sentence by treating *it* as a *noun* but cannot obtain *voltmeter* as the subject. Moreover, CCG completely is not able to parse pronouns of other forms (e.g., them, its, their, that, which) and struggles to represent knowledge when pronouns represent indexicality. The anaphora depends on lexical semantics and its resolve depends on the reasoning through the ontology of the prototype.

CCG is unable to parse sentences that start and end with a prepositional phrase [23]. For example, *in series circuit, the current is a single current*- this sentence is not parsed by CCG. In contrast, *the current is a single current in series circuit*- is sometimes parsed by CCG. The lexicon the prototype is using has nine different types of prepositions. Sometimes, it is difficult to even identify regular prepositions. For instance, *the sum of potential differences in a circuit adds up to zero voltage*- Though in regular grammars, *up* is not treated as adverbs- these are called particles where prepositions have no objects and require specific verbs with them (e.g., throw out, add up). The parsing ability of the prototype increased as we defined grammar rules for such prepositions.

*Complementizer*, although it is a form of preposition, it is not recognized by CCG. *Adverbs*, on the other hand, have a strong coverage by CCG. In many cases, adverbs sit at the end of the sentence- CCG does not provide any category to define these adverbs although it has fully featured adverb categories for other two positions of an adverb in sentence- adverbs that start a sentence or that sit in the middle of a sentence. These issues have been resolved by adding new grammar rules.

The lexicon has two categories for *coordinators*- sitting at the beginning of a sentence (e.g., since, as) and relating two clauses (e.g., and, or). CCG defined that they can be in the middle of two noun phrases only with *np\np/np* but the sentence *series and parallel circuits are the types of circuits* has the category *n\n/n* rather than *np\np/np*. CCG handles adverbs and conjunctions well but it seriously lags in handling sentences having similar verbs as in the sentence *the sum of current flowing into the junction is eventually equal to the sum of current flowing out of the junction*. The identical verbs *flowing (gerund)* appear twice with another *verb (be)* is concerning. Moreover, a verb has to be present in a sentence to form predicate argument structure but we discovered that there are sentences which do not have any verbs- *the bigger the resistance, the smaller the current.* Gerund of verb is known as noun. *Gerund* is formed by placing *ing* at the end of the verb. For example, *current flowing into a junction is equal to the current flowing out of the junction*- in this sentence, *flowing* is a gerund. Gerunds are not treated as nouns in CCG. In other words, gerunds, if treated as nouns in CCG, the sentence struggles to be parsed.

After creating grammar rules and phrasal structures and adding them into the lexicon and morphology of the prototype, the parsing ability of the prototype increased to 31 percent (Table IV). Although the prototype was tested with a limited dataset, it was unable to parse any sentence from the corpus before the evaluation.

TABLE IV
PARSING ABILITY OF THE TKM PROTOTYPE

| State of the Prototype | Total Sentences | Parsed Sentences | Efficiency |
|---|---|---|---|
| Preliminary | 1,029 | 0 | 0% |
| Evaluated | 981 | 300 | 31% |

We analyzed the 300 sentences parsed by the prototype and figured out the number of subject, object and verb they consist. In Fig. 5 we see that the prototype works well when the number of subjects and objects in a sentence do not exceed two and when the number of verbs does not exceed one.

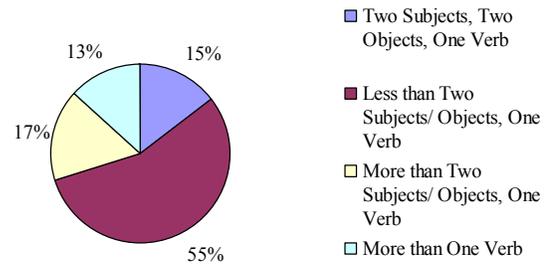

Fig. 5. Number of subjects, objects and verbs in the sentences parsed by the prototype

The inefficiency of the prototype to parse sentence is due to the absence of phrasal structures (hence, the categories). 69 percent of the sentences in the corpus have phrasal structures that are not supported by the CCG structure. It should be noted that the prototype fails to parse sentences even for absence of just one category. For example, *One simple DC circuit consists of a voltage source (battery or voltaic cell) connected to a resistor* – this sentence is not parsed by the prototype for the absence of category of conjunction *or* (np\n/np) and for the category of verb *connected* (s\np/pp). In the corpus, these absent categories are identified so that modification of the lexicon becomes easier.

V. CONCLUSION

In this paper, we represented a corpus-based evaluation of lexical components of a domain-specific NLP system. We proposed a stochastic approach to develop a domain-specific representative corpus. The linguistic resources of the corpus have been used to evaluate the lexical components of the system. As knowledge representation is the goal of the system and successful parsing and information retrieval is at the heart of knowledge modelling, we adjusted the lexical components of the system to increase its parsing ability. This evaluation made the TKM prototype an ideal knowledge representation tool for the domain DC electrical circuit as it parses large number of text from the domain and the information retrieval process has been satisfactory. However, the evaluation on the knowledge model and knowledge representation depends on the human cognition of the domain and its representation based on the semantic and linguistic relations in text. We developed a conceptual structure and a framework of semantic relations for the domain [24] and the evaluation of its knowledge model will be accomplished once we are able to tie the significance of linguistic relations in the text with the conceptual structure for the domain.

REFERENCES

[1] W. Ou and A. Elsayed, "A Knowledge-Based Approach for Semantic Extraction", *International Conference on Business Knowledge Management*, Macao, 2006.

[2] T. Declerck, J. Klein, and G. Neumann, "Evaluation of the NLP Components of an Information Extraction System for German", *Proceedings of the first international Conference on Language Resources and Evaluation (LREC)*, Granada, 1998, pp. 293-297.

[3] D. Evans, "Corpus building and investigation for the humanities",Available: http://www.humcorp.bham.ac.uk/humcorp/information/corpusintro/Unit1.pdf [27 September 2008]

[4] K. D. Forbus, "Qualitative process theory: Twelve years after", Artificial Intelligence, volume 59, number 1-2, 1993, pp. 115-123.

[5] M. Palmer and T. Finin, "Workshop on the evaluation of natural language processing systems", Computational Linguistics, volume 16, number 3, 1993, pp. 175-181.

[6] T. McEnery and A. Wilson, Corpus Linguistics, Edinburgh: Edinburgh University Press, United Kingdom, 1996.

[7] J. Klein, T. Declerck, and G. Neumann, "Evaluation of the Syntactic Analysis Component of an Information Extraction System for German", *Proceedings of the 1st International Conference on Language Resources and Evaluation*, Granada, Spain, 1998, pp. 293-297.

[8] G. Neumann, R. Backofen, J. Baur, M. Becker, and C. Braun, "An Information Extraction Core System for Real World German Text Processing", *Proceedings of the 5th Conference on Applied Natural Language Processing (ANLP)*, USA, 1997, pp. 209-216.

[9] G. Grefenstette, "The WWW as a Resource for Example-based MT Tasks", *Proceedings of ASLIB Translating and the Computer Conference*, London, United Kingdom, 1999.

[10] M. Volk, "Exploiting the WWW as a Corpus to Resolve PP Attachment Ambiguities", *Proceedings of the Corpus Linguistics*, Lancaster, United Kingdom. 2001.

[11] N. Modjeska, K. Markert, and M. Nissim, "Using the Web in Machine Learning for Other-anaphora Resolution", *Proceedings of EMNLP*, Sapporo, Japan, 2003, pp. 176-183.

[12] S. Clark, M. Steedman, and J. R. Curran, "Object-Extraction and Question-Parsing using CCG", Available: http://www.iccs.inf.ed.ac.uk/~stevec/papers/emnlp04.pdf [27 September 2008]

[13] W. Finkler and G. Neumann, "Morphix: A fast Realization of a Classification–based Approach to Morphology", *Proceedings der 4. Österreichischen Artificial Intelligence Tagung, Wiener Workshop Wissensbasierte Sprachverarbeitung*, Berlin, 1988, pp. 11-19.

[14] V. Liu and J. Curran, "Web Text Corpus for Natural Language Processing", *Language Technology Seminar Series*, Sidney, Australia, 2006.

[15] A. Kilgarriff and G. Grefenstette, "Introduction to the special issue on the web as corpus", Computational Linguistics, volume 29, 2006, p. 333–347

[16] M. Baroni and S. Bernardini, "BootCaT: Bootstrapping Corpora and Terms from the Web", *Proceedings of LREC*, 2004.

[17] M. Baroni, A. Kilgarriff, J. Pomikálek, and P. Rychlý, "WebBootCaT: a web tool for instant corpora", *Proceeding of the EuraLex Conference*, Italy, 2006, pp. 123-132.

[18] T. McEnery, R. Xia, and Y. Tono, Corpus-Based Language Studies: An Advanced Resource Book, London: Routledge, 2006.

[19] D. Biber, "Representativeness in Corpus Design", Literary and Linguistic Computing, Volume 8, Number 4, 1993, pp. 243-257.

[20] University of Illinois at Urbana-Champaign, "The SNoW Learning Architecture", Available: http://l2r.cs.uiuc.edu/~danr/snow.html [27 September 2008]

[21] A. S. Hornby, Guide to Patterns and Usage in English, 2nd Edition, Oxford University Press, Delhi, 1995, pp. 1-2.

[22] M. A. Covington, Natural Language Processing for Prolog Programmers, 1st Edition, Prentice Hall, 1993, pp. 88-90.

[23] M. Steedman and J. Baldridge, "Combinatory Categorial Grammar", Unpublished Tutorial Paper, 1993.

[24] R. Shams and A. Elsayed, "Development of a Conceptual Structure for a Domain-specific Corpus", *3rd International Conference on Concept Maps (CMC)*, Estonia and Finland, 2008.